\documentstyle[12pt,tighten,aps,a4,graphicx,amssymb,floats]{revtex}

\begin{document}

 \title{Hard Disks in Narrow Channels}

 \author{Ch. Forster}
 \address{Institut f\"ur Experimentalphysik, Universit\"at Wien,
 Boltzmanngasse 5, A-1090 Wien, Austria \\ E-mail: tina@ap.univie.ac.at}

 \author{D. Mukamel}
 \address{Department pf physics of complex systems,
 Weizmann Institute if Science\\
 Rehovot, Israel 76100 \\ E-mail:david.mukamel@weizmann.ac.il }

 \author{H. A. Posch}
 \address{Institut f\"ur Experimentalphysik, Universit\"at Wien,
 Boltzmanngasse 5, A-1090 Wien, Austria \\ E-mail: posch@ls.exp.univie.ac.at}

 \date{\today}

 \maketitle

\begin{abstract}
   The thermodynamic and dynamical behavior of a gas of hard disks
in a narrow channel is studied theoretically and numerically.
Using a virial expansion we find that the pressure and collision
frequency curves exhibit a singularity at a channel width
corresponding to twice the disk diameter. As expected, the maximum
Lyapunov exponent is also found to display a similar behavior. At
high density these curves are dominated by solid-like
configurations which are different from the bulk ones, due to the
channel boundary conditions.
\end{abstract}

\section{Introduction}
\label{Introduction}
     The thermodynamic behavior of hard-disk systems has been a subject of
considerable interest in recent years \cite{Stillinger}. In
particular, the freezing transition taking place in these systems
has been analysed by computer simulations for cases where the
aspect ratio of the simulation box is close to one. Furthermore,
the dynamical stability of such systems has been studied in detail
and Lyapunov spectra have been computed\cite{PH00,FHPH04}. The
maximum Lyapunov exponent was found to exhibit a maximum in the
density regime characteristic for the fluid-to-solid
transition\cite{DPH96}. In a restricted geometry, such as a narrow
channel, one expects other interesting features to emerge when the
width of the channel becomes comparable to the disk's size.
Studying narrow channels also enables one to examine
long-wavelength perturbed states associated with the small
positive Lyapunpov exponents found in the tangent-space dynamics
of hard-disk and hard-sphere systems \cite{PH00,FHPH04}. This
provided the motivation for us to re-investigate the thermodynamic
and dynamical properties of a hard-disk gas in a narrow channel.

    It is found that for narrow channels the pressure exhibits a singularity
at a channel width equal to twice the diameter of the disks.
Narrower channels do not allow particles to pass each other. At
the same width a singularity is also observed for the collision
frequency. In addition, the maximum Lyapunov exponent and the
Kolmogorov-Sinai entropy display a non-trivial dependence on the
channel width in this regime.

   The model we study consists of $N$ hard disks in a two-dimensional
box with side lengths $L_x,L_y$ and aspect ratio $A=L_y/L_x$. We
vary the width of the box, keeping the volume $V = L_x L_y$ and
the particle density $\rho = N/V$ constant.  Very small aspect
ratios $A \ll 1$  are considered, for which the box resembles a
narrow channel. For most of our work we use {\em periodic}
boundary conditions in both directions. Exact thermodynamic
properties are known for extremely narrow channels in this case
$L_y < \sqrt{3} \sigma$ \cite{Woj82,Woj83}. Recently, the transport
properties of two particles in
a square periodic box, $A = 1$, have been studied by computer
simulation \cite{VG03}. The case of two particles with {\em reflecting}
boundary conditions in both directions has also received
attention, both by molecular dynamics \cite{Awazu} and analytical
approaches \cite{Munakata}. It was found that for certain aspect
ratios the compressibility becomes negative. We find
analogous results also for partly-reflecting boundaries, which are reflecting
for the walls parallel to the long direction of the box, $x$, and periodic
for the walls parallel to $y$.

    Throughout, reduced units are used for which
the disk diameter $\sigma$, the particle mass $m$ and the kinetic
energy per particle, $K/N$, are unity. There is no potential
energy in this case, and the total energy $E$ is identical to the
kinetic energy $K$. As a consequence, the temperature $T$ is just
an irrelevant parameter and is fixed to unity. Also Boltzmann's
constant $k$ is taken unity.

    The paper is organized as follows. In Section \ref{theory} we consider the
low-density limit. In Section \ref{EOS} we present an analytical
derivation of the pressure and show that the singularity taking
place at $L_y = 2$ already exists in this limit. In addition, we
obtain an analytical expression for the collision frequency which
shows a behavior very similar to that of the pressure. Results of
numerical simulations of these quantities and of the Lyapunov
spectra are presented in Section \ref{NS}. A numerical study of
the high-density case is presented in Section \ref{HD}.
Full Lyapunov spectra for periodic systems with periodic boundaries are
discussed in Section \ref{spectra}.
We close with a brief discussion of the results in Section
\ref{discussion}.

\section{Low-density limit for systems with periodic boundaries}
\label{theory}
\subsection{Virial approach to the equation of state and collision
          frequency }
\label{EOS}
    The pressure in the low-density limit is obtained  by
the leading terms in the virial expansion as applied to the
narrow-channel geometry. Let $q(L_y)$ be the excluded volume of a
single disk in the channel. The phase-space volume associated with
placing $N$ disks in the channel is given by
\begin{equation}
\Omega(N) = V(V-q) \dots (V-(N-1)q) ,
\end{equation}
which, to leading order in the density, is reduced to
\begin{equation}
\Omega(N) \simeq V^N \left( 1 - \frac{N(N-1)}{2} \frac{q}{V}\right)
        \simeq V^N \left(1 -\frac{q}{2 v}\right)^{N-1} .
\end{equation}
Here,  $v=V/N$ is the specific volume.
For low density the specific  entropy of the gas is
given by
\begin{equation}
s \equiv \frac{S}{N}=  k \ln \left(v-{q(L_y)\over 2}\right),
\label{entropy}
\end{equation}
where $k$ is Boltzmann's constant.

For $L_y > 2$ the excluded area  is simply given by the area of a
disk of radius $1$, namely $q(L_y>2)=\pi$. To leading order in the
density, the pressure $P$ is given by
\begin{equation}
{Pv \over kT}=v {\partial s \over \partial v}\simeq 1+{\pi \over
2v}~,
\label{compressionfactor}
\end{equation}
and is thus independent of the channel width $L_y$. This is the
expected leading-order expression of the virial expansion of a gas
of hard disks.

\begin{figure}
\centering{\includegraphics[width=8cm]{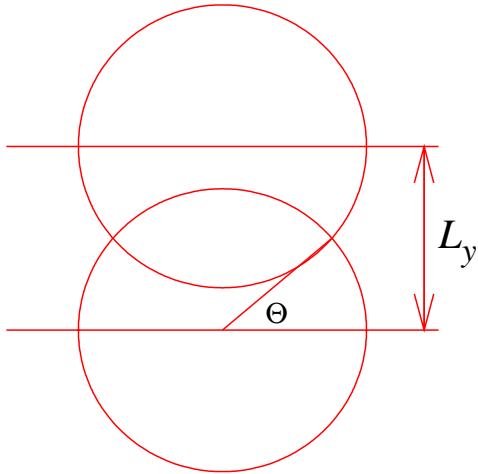}} \vspace{3mm}
\caption{The excluded area associated with a disk in a channel of
width $L_y<2$ for the case of periodic boundary conditions. The
area $q(L_y)$ is given by that of the union of the disk and its
translationally-displaced image within the channel.}
\label{geometry}
\end{figure}

For $1<L_y<2$, however,  the
excluded area is reduced due to the interference with the disk
image resulting from the periodic boundary condition in the $y$
direction (see Fig. \ref{geometry}). Simple geometrical
considerations yield
%
\begin{equation}
q(L_y)=2\theta + \sin(2\theta) ,
\end{equation}
where
\begin{equation}
L_y=2 \sin(\theta) .
\end{equation}
The excluded volume $q(L_y)$ depends on $L_y$. Taking the
derivative of the entropy with respect to the volume and keeping
the aspect ratio $A\equiv L_y/L_x$ constant, one finds to leading
order in the density
\begin{equation}
{Pv \over kT}=1+{q(L_y) \over 2v}-{L_y \over 4v}\sqrt{4-L_y^2}~.
\label{pvkt}
\end{equation}
Note that the pressure curve exhibits a square root
singularity as one approaches $L_y = 2$ from below.

    The singularity at $L_y = 2$ does not signify a phase transition in
the usual sense as it is not a  collective phenomenon. In fact,
the singularity exists even for finite $N$ and it explicitly shows
up in the second term of the virial expansion. This is different
from the usual liquid-solid transition in the bulk, which is
obtained only after summing over all orders of that expansion. The
singularity in the narrow channel is a consequence of the fact
that the available volume for the disks is a singular function of
the control parameter $L_y$ at $L_y = 2$. Higher order terms in
the virial expansion are expected to exhibit a singularity at
other integer channel widths, $L_y = 3,4...$ . Thus singularities
in the pressure are expected to show up at these widths at higher
densities.

    We now turn to the evaluation of the collision frequency $\nu_2$.
The virial expression for the pressure, appropriately modified for
two-dimensional hard disks, is given by \cite{HA67}
\begin{equation}
\frac{P V}{N k T} = 1 + \frac{1}{2NkT \tau} \sum_{c} {\bf r}_c
            \cdot {\bf v}_c \; ,
\label{virial}
\end{equation}
where the sum is over all collisions taking place during the time
interval $\tau$. Here, the relative position vector of two colliding
particles $i$ and $j$ at the time of the collision, ${\bf r}_c
\equiv {\bf r}_i -  {\bf r}_j$, satisfies $|{\bf r}_c| = \sigma$,
and ${\bf v}_c \equiv {\bf v}_i -  {\bf v}_j$ is the corresponding
relative pre-collision velocity. This may be written as
\begin{equation}
\frac{PV}{NkT} -1 = \frac{1}{2 k T} \nu_2 \Delta p
\label{delp}
\end{equation}
where $ \Delta p$ is the average momentum transfer in a collision.
For a large aspect ratio $A$ the collision frequency is given by
\cite{ZvBD}
\begin{equation}
\nu_2 = 2 \pi^{1/2} (kT)^{1/2} (\sigma/v) g(\sigma) ,
\label{nu2}
\end{equation}
where $g(r)$ is the pair distribution function. At contact, $r = \sigma$,
and in the low-density limit one has  $g(\sigma) = 1$.
Combining this result with the Eqs. (\ref{compressionfactor}) and (\ref{delp}),
one finds for the average momentum transfer
\begin{equation}
    \Delta p = (\sqrt{\pi} / 2 ) \sqrt{kT}  .
\label{deltap}
\end{equation}
This expression may also be derived directly from the
Maxwell-Boltzmann velocity distribution. It is expected to hold also
for narrow channels. The reason is that in the narrow channel the
velocity distribution is still isotropic and is given by the
Maxwell-Boltzmann distribution (see Section \ref{NS}). Therefore,
one can use Eq. (\ref{deltap}), together with Eq. (\ref{delp}), to
obtain the collision frequency for narrow channels:
\begin{equation}
  \frac{ \nu_2} {\sqrt{kT}} = \frac{4}{\sqrt{\pi}}
            \left( \frac{Pv}{kT} -1\right)
\label{cf}
\end{equation}
This expression is valid both below and above $L_y = 2$.

     It is expected that in such a highly anisotropic system the
pressure tensor is not isotropic. In the low-density limit one can
evaluate the diagonal components of the pressure tensor, $P_{xx}$
and $P_{yy}$,  using again the expression for the entropy in Eq.
(\ref{entropy}).  One finds
\begin{eqnarray}
P_{xx} & = & L_x \left(\frac{\partial s}{\partial L_x}\right)_{L_y}
     = \frac{v} {v - (1/2) q(L_y)},  \nonumber \\
P_{yy} & = & L_y \left(\frac{\partial s}{\partial L_y}\right)_{L_x}
     = \frac{v - (1/2) L_y \sqrt{4 - L_y^2}}{v - (1/2) q}. \nonumber \\
\end{eqnarray}
Furthermore, one has
$$
   v \left(\frac{\partial s}{\partial v}\right)_A  =
  \frac{1}{2} L_x \left(\frac{\partial s}{\partial L_x}\right)_{L_y} +
  \frac{1}{2} L_y \left(\frac{\partial s}{\partial L_y}\right)_{L_x},
$$
and, hence, $P = (P_{xx} + P_{yy})/2$.

   It is evident that in the low-density limit the pressure
tensor is isotropic for $L_y > 2$. It becomes anisotropic for $L_y
< 2$. At higher densities the two pressure-tensor components
differ even above $L_y = 2$ (see Section \ref{HD}). The
square-root singularity of the total pressure $P$ at $L_y = 2$
originates from $P_{yy}$. The other component, $P_{xx}$, exhibits
a weaker singularity.

\subsection{Numerical simulations}
\label{NS}
Here we carry out numerical simulations for the computaton of the
pressure, the collision frequency and the Lyapunov exponents. The
pressure is evaluated from the impulsive version of the virial
theorem Eq. (\ref{virial}) \cite{Rapaport,AT}. The collision
frequency of a single particle, $\nu_2$, is obtained from the
total number of collisions per unit time, divided by $N/2$.  For
the Lyapunov spectrum we use the algorithm outlined in Refs.
\cite{DPH96} and \cite{PH00}.  For the {\em periodic}-boundary case the
linear momentum is conserved both in $x$ and $y$ directions, and,
as expected, altogether 6 Lyapunov exponents vanish due to the
conservation of energy, momentum, center of mass (only in tangent
space), and the regularity of the dynamics in the phase-flow
direction. For the {\em partly-reflecting} boundary case only four of the
exponents vanish due to a lack of momentum and center-of-mass
conservation in the $y$ direction. The Kolmogorov-Sinai (KS)
entropy, $h_{KS}$, which measures the rate of information gain on
the initial conditions due to the time-reversible dynamics, is given by
the sum of
the positive exponents\cite{Oseledec}. Since its dependence on the
channel width is qualitatively similar to that  of the maximum
Lyapunov exponent $\lambda_1$ we concentrate on the latter in the
following.

\begin{figure}
\centering{\includegraphics[width=10cm]{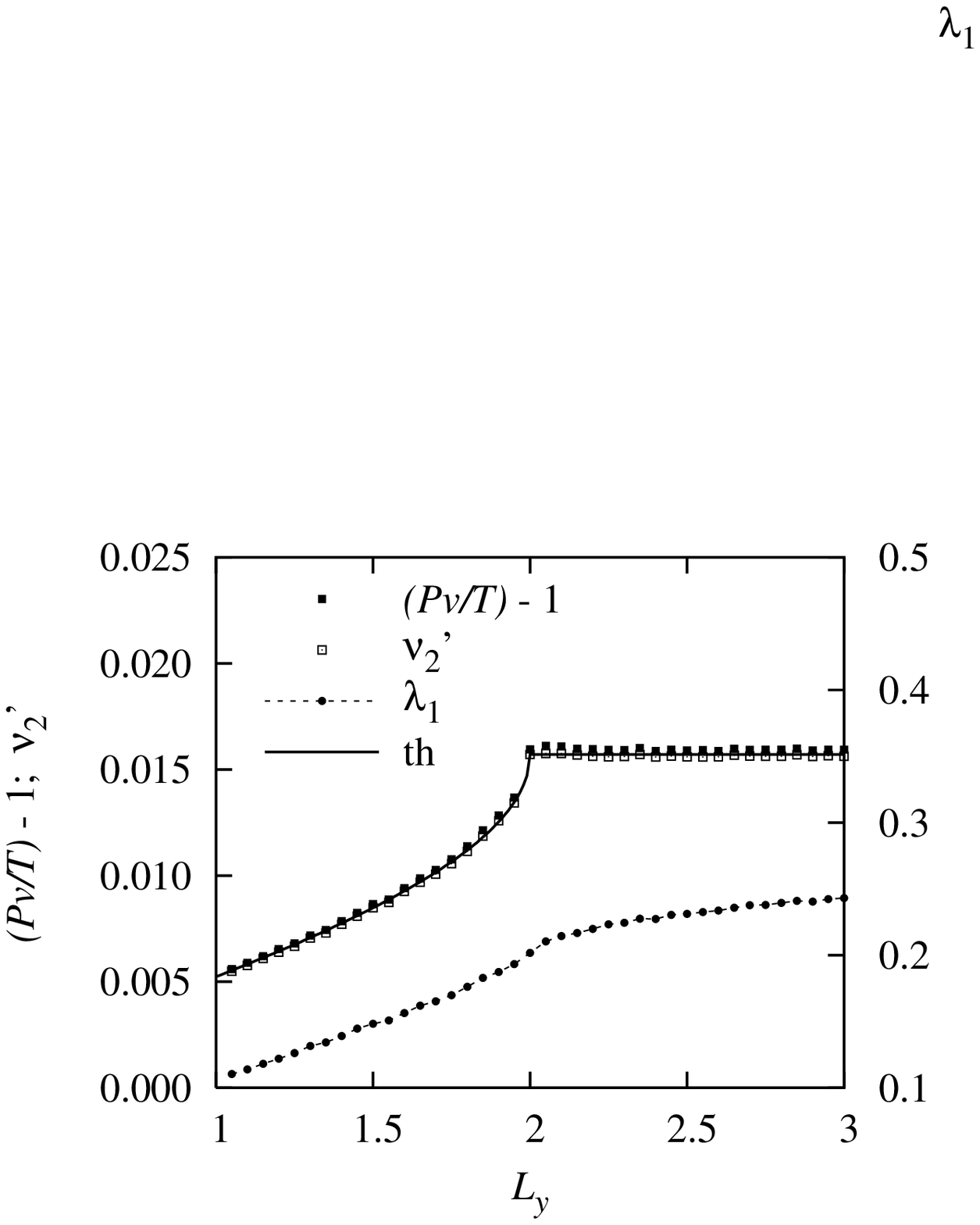}}
\caption{Theoretical and simulation results for $N = 20$ particles and
a density $\rho = 0.01$
for the periodic-boundary case. Shown is the dependence of the maximum
Lyapunov exponent $\lambda_1$ (right vertical scale), and of the
equation of state $(Pv/T) - 1$ (left vertical scale, full squares), and the
rescaled single-particle collision
frequency $\nu_2' = (\sqrt{\pi}/4) \nu_2$ (left vertical scale, open squares)
on the channel width $L_y$. The full line is computed from
Eq. (\ref{pvkt}).}
\label{Fig1}
\end{figure}

   In Figure \ref{Fig1} we compare the theoretical curves for the
pressure and collision frequency, Eqs. (\ref{compressionfactor}),
(\ref{pvkt}) and (\ref{cf}), with the numerical results. In this
low-density regime an excellent agreement  is obtained. Also shown
in this figures is the channel-width dependence of the maximum
Lyapunov exponent. It shows a shoulder at the critical width $L_y
= 2$.
The tangent-space perturbation associated with $\lambda_1$ has been
shown to be localized in space \cite{FHPH04}.
This exponent is closely connected
to the collision frequency and, as a consequence, has a similar
qualitative behavior. This is particularly pronounced in the high
density regime as dealt with in the next section.

\begin{figure}
\centering{\includegraphics[width=10cm]{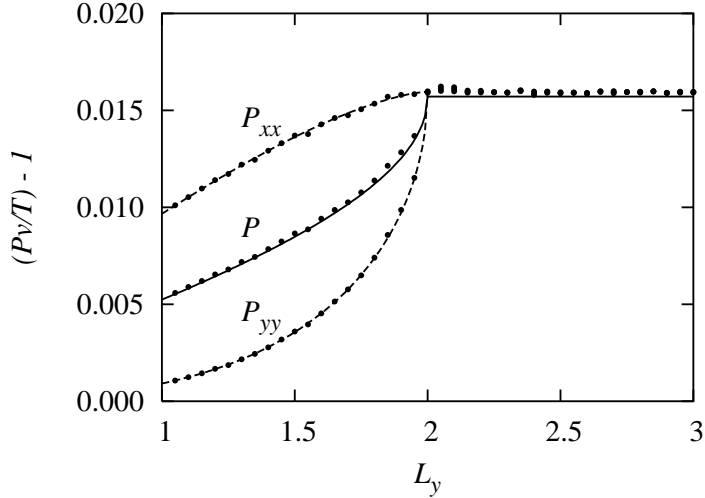}}
\caption{ Comparison of the $L_y$-dependence of the theoretical
curves for $(P v/T) - 1$,  $(P_{xx} v/T) - 1$, and
$(P_{yy} v/T) - 1$ (lines)
with numerical simulations (points) for a system with 20 disks at a
density $\rho = 0.01$.}
\label{Pxx_20_0.01}
\end{figure}

    In Fig. \ref{Pxx_20_0.01} we compare the theoretical curves for the
pressure-tensor components $P_{xx}$ and $P_{yy}$ with computer
simulations for a density $\rho = 0.01$ and find excellent
agreement. As noted above, it is evident that the square-root
singularity of $P$ originates from $P_{yy}$.

\begin{figure}
\centering{\includegraphics[width=10cm]{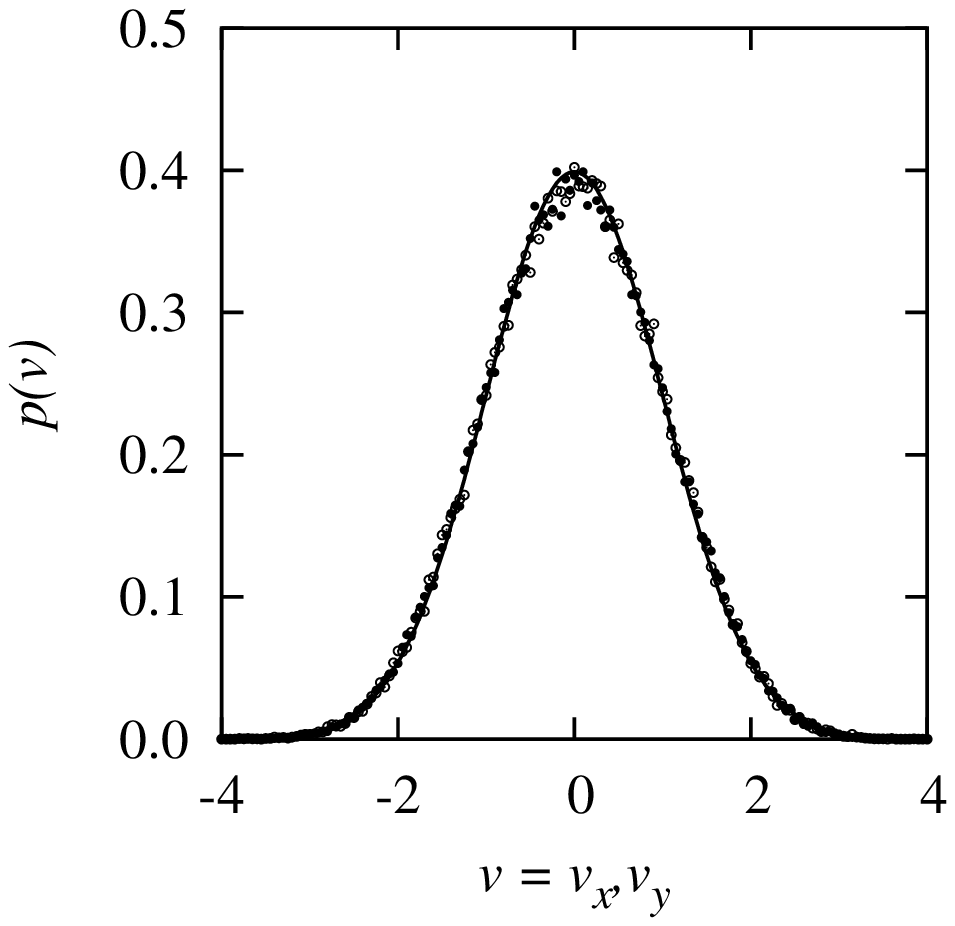}}
\caption{Distribution of the $x$ and $y$ components of the
velocity (open and full circles) for a box of width $L_y = 1.8$.
The full line corresponds to a Gaussian distribution.}
\label{Fig2}
\end{figure}

    It is interesting to note that in the narrow-channel regime,
$L_y < 2$, the velocity distribution is still isotropic and is
given by the Maxwell-Boltzmann distribution as is shown in Fig.
\ref{Fig2}. This supports the use of Eq. (\ref{deltap}) for the
average momentum transfer at a collision even for $L_y < 2$.

    The singular behavior of the pressure is due to a sharp
increase of the collision frequency as one approaches $L_y = 2$
from below. As can be seen in Fig. \ref{Fig1} the maximum Lyapunov
exponent is reminiscent of this singularity as it is closely
related to the collision frequency. At this low density this
singularity is not clearly pronounced. However as will be shown
below, at high densities the existence of a cusp at $L_y = 2$ is
evident. The increase of the collision frequency is due to
arresting configurations in which a pair of disks become trapped
due to the boundaries.

\section{High density regime}
\label{HD}

\begin{figure}[h]
\centering{\includegraphics[width=10cm]{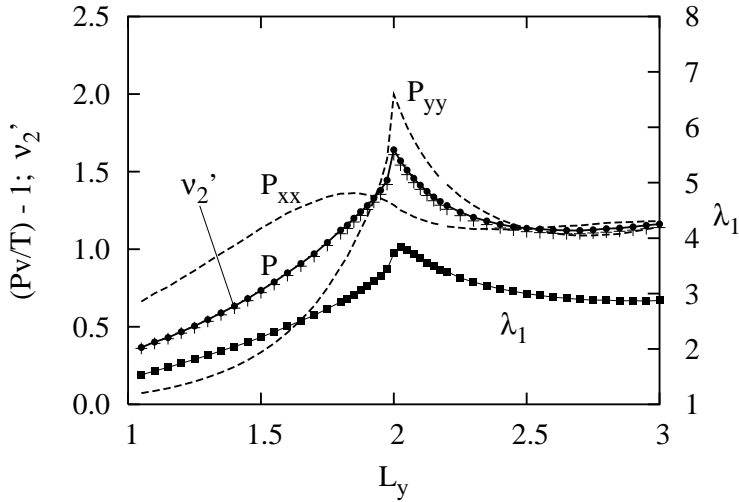}}
\caption{Simulation results for $N = 20$ particles and a density $N/V=0.4$
for the periodic-boundary case. Shown is the dependence of the maximum
Lyapunov exponent, $\lambda_1$ (right vertical scale), of the
equation-of-state function $(P v/kT) - 1$ (full circles with label $P$,
left vertical scale), and of the rescaled single-particle collision
frequency, $\nu_2' = \left[(\pi/kT)^{1/2}/4\right] \nu_2$
(crosses, left vertical scale)
on the channel width $L_y$. $\nu_2'$ agrees well with $P$. The dashed
curves labelled $P_{xx}$ and
$P_{yy}$ give the contributions  $(P_{xx} v/kT) - 1$ and
$(P_{yy} v/kT) - 1$ of the respective pressure-tensor components.}
\label{Fig_0.4}
\end{figure}

  For hard-disk systems at high density one has to rely only on numerical
simulations. In Fig. \ref{Fig_0.4} we present the pressure, the collision
frquency and the maximum Lyapunov exponent as a function of $L_y$ at a
density $\rho = 0.4$.
It is observed that the singularity taking place at $L_y = 2$
persists at high density in all three curves. According to Eq.
(\ref{cf}) the curves for the collision freuqency and for $Pv/T -
1$ are proportional with a propertionailty constant independent of
the density. This is evidently confirmed in Fig. \ref{Fig_0.4}.
For the special case of only $N = 2$ particles in a square
periodic box, the singularity at $L_y = 2$ appears at the critical
density $ \rho_c = 0.5 $ and has recently been observed
\cite{VG03}.

\begin{figure}
\centering{\includegraphics[width=10cm]{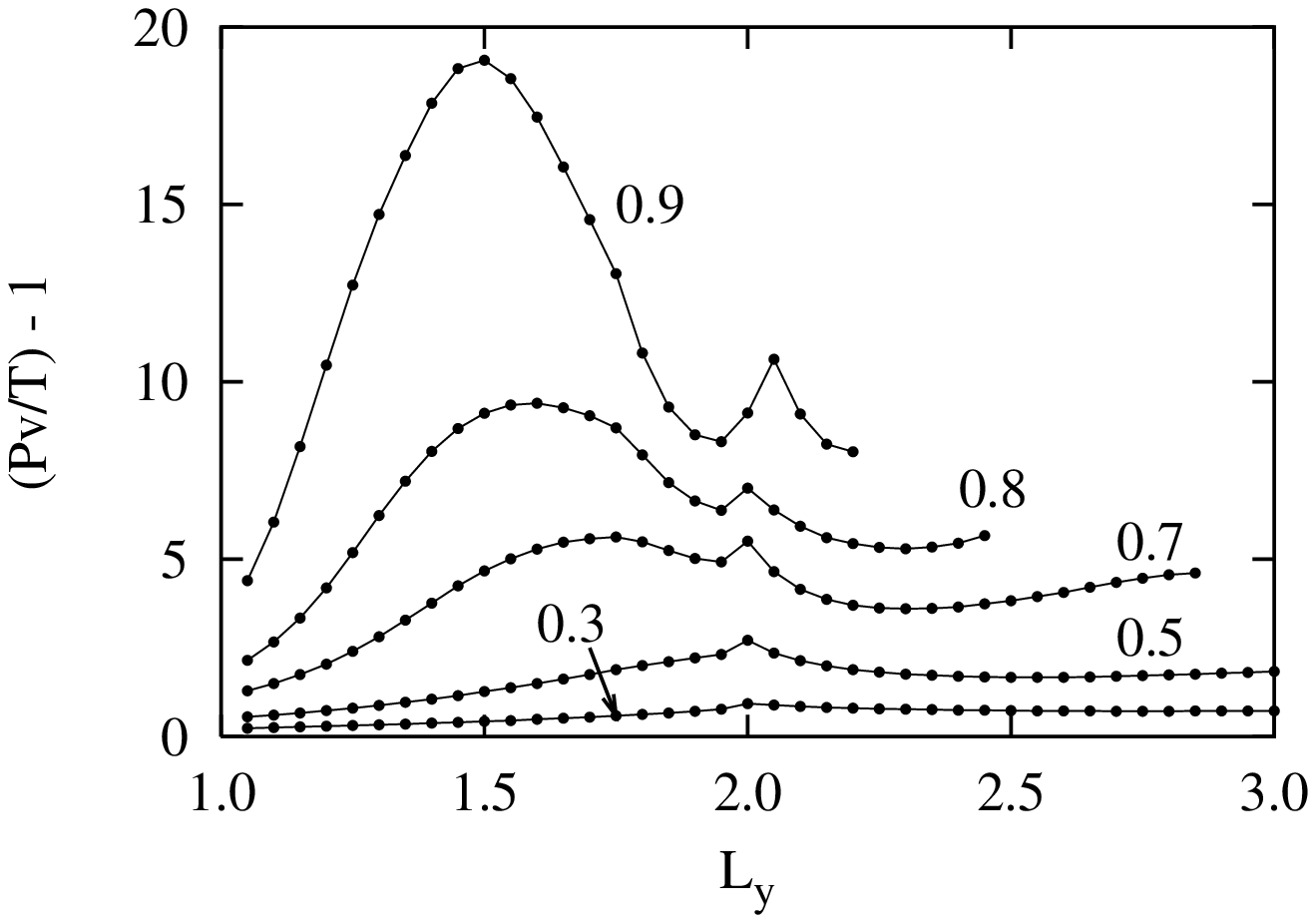}}
\caption{$L_y$ dependence of the pressure for various densities
as indicated by the labels.}
\label{many}
\end{figure}

\begin{figure}
\centering{\includegraphics[width=10cm]{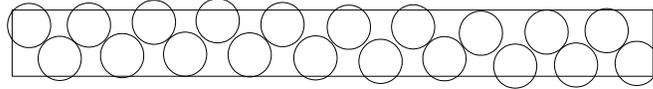}}
\caption{A typical microscopic configuration for a density $\rho = 0.9$
and a channel width $L_y = 1.5$. It resembles the closest-dense packing
consistent with the periodic boundaries. It minimizes the mean free path
and, hence, maximizes the  collision frequency and the pressure.}
\label{structure_0.9}
\end{figure}

    When the density is increased beyond 0.5, a broad peak
emerges at some $ L_y < 2 $ in the curves for the pressure and the
maximum Lyapunov exponent. This is demonstrated in
Fig. \ref{many}. This feature is attributed to the
pronounced short-range crystalline-like order which is induced by
the narrow-channel boundary conditions. Note that this order is not the
natural triangular lattice of the system in the bulk. A typical microscopic
configuration corresponding to this structure for a density 0.9 and
$L_y = 1.5$ is shown in Fig. \ref{structure_0.9}. The particles
are arrested and cannot travel across the system, which has been
referred to as the localized regime \cite{VG03}.

\begin{figure}
\centering{\includegraphics[width=10cm]{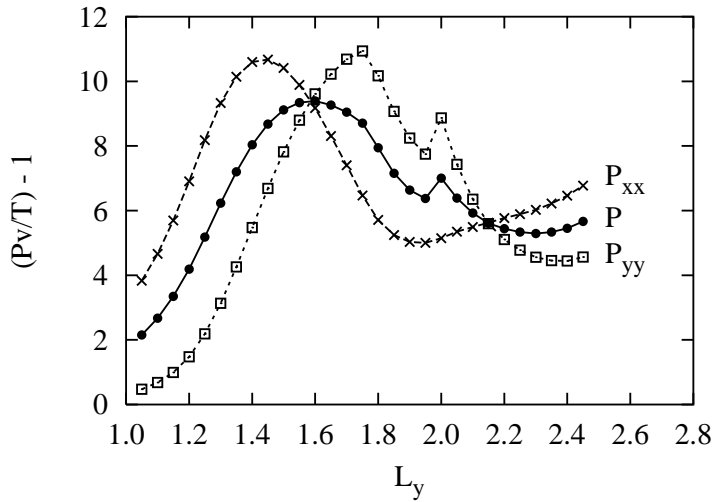}}
\caption{Simulation results for the pressure curves
$(Pv/T)-1$, $(P_{xx} v/T)-1$, and $(P_{xx} v/T)-1$ as a function of the
channel width $L_y$. The system consists of $N = 20$ particles at
a density $\rho = 0.8$. }
\label{Pxx_20_0.8}
\end{figure}

    We have computed also the two pressure-tensor components
$P_{xx}$ and $P_{yy}$ for a system with a density $\rho = 0.8$.
As can be seen in Fig. \ref{Pxx_20_0.8}, the two components are
significantly different both above and below $\L_y = 2$. The
singularity of the pressure at $L_y = 2$ is evidently caused by
$P_{yy}$, as is also the case for low densities.

\begin{figure}
\centering{\includegraphics[width=10cm]{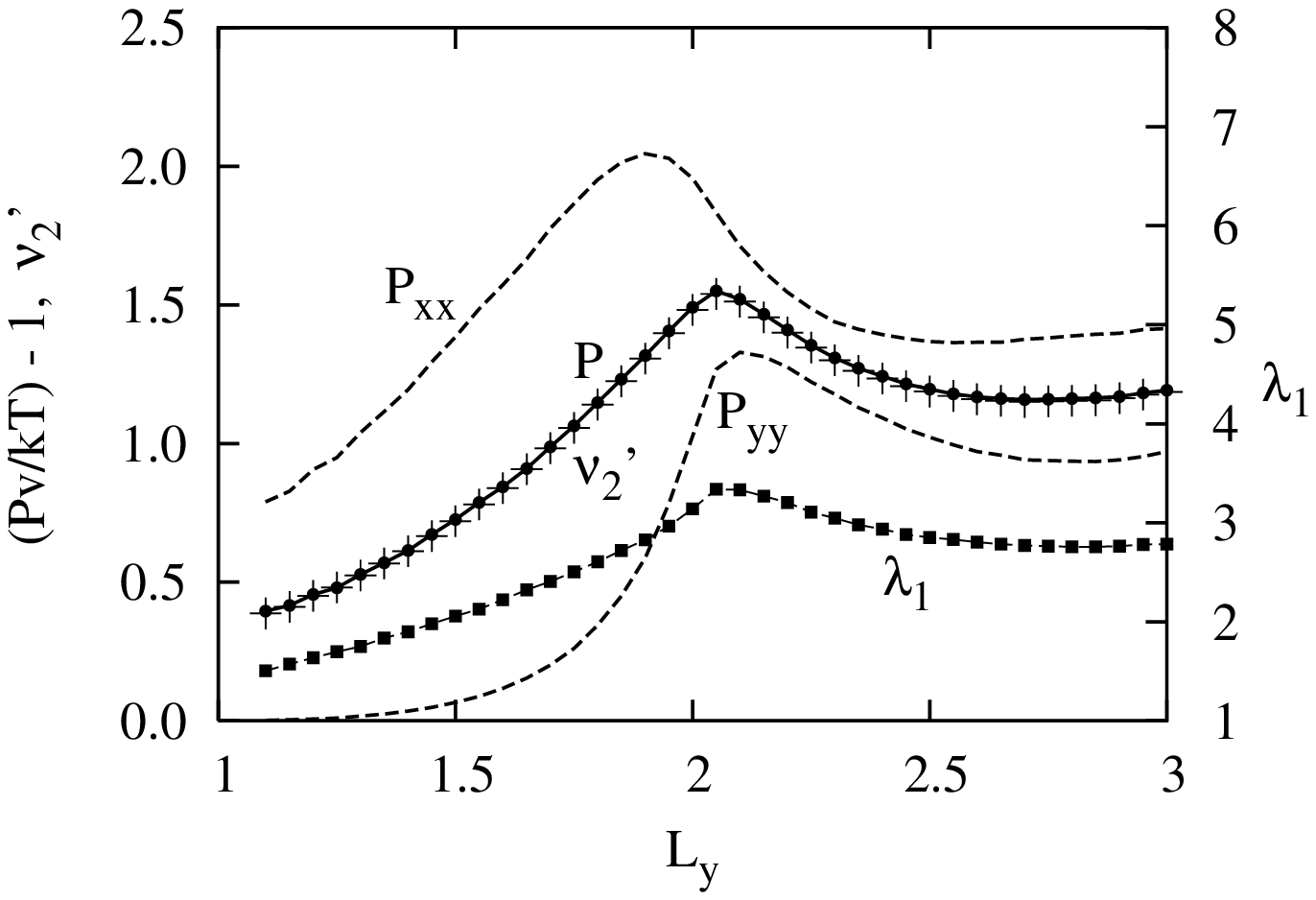}}
\caption{Simulation results for $N = 20$ particles and a density $N/V=0.4$
in a rectangular box with partly-reflecting boundaries (see the main text).
Shown is the dependence of the maximum
Lyapunov exponent, $\lambda_1$ (right vertical scale), of the
equation-of-state function $(P v/kT) - 1$ (full circles with label $P$,
left vertical scale), and of the rescaled single-particle collision
frequency, $\nu_2' = \left[(\pi/kT)^{1/2}/4 \right]\nu_2$
(crosses, left vertical scale) on the channel width $L_y$.
$\nu_2'$ agrees well with $P$. The dashed curves labelled $P_{xx}$ and
$P_{yy}$ give the contributions  $(P_{xx} v/kT) - 1$ and
$(P_{yy} v/kT) - 1$ of the respective pressure-tensor components.}
\label{Fig4}
\end{figure}

    All the examples so far are for the case of periodic boundaries
in both the $x$ and $y$ directions. A qualitatively similar result
is obtained, if the long sides of the channel are elastically
reflective, whereas the short sides remain periodic. This is
demonstrated in Fig. \ref{Fig4} for $N = 100$ particles  at a
density of $N/V = 0.4$, where $P$,  $\lambda_1$ and $\nu$ are
shown as a function of the box width $L_y$. The maxima are not as
sharp as in Fig. \ref{Fig_0.4} for the periodic case, but are as
pronounced.

    In order to get some insight into the dynamics of such systems it
is usually useful to study the diffusion coefficient. It is well known that
this coefficient does not exist in two-dimensional
systems in the thermodynamic limit \cite{AW70}. However, it is possible to
compute the mean-squared
displacement for a {\em finite} system and to extract effective diffusion
constants $D_x$ and $D_y$ in $x$ and $y$ directions, respectively.
They are obtained from fits to the
\begin{figure}
\centering{\includegraphics[width=10cm]{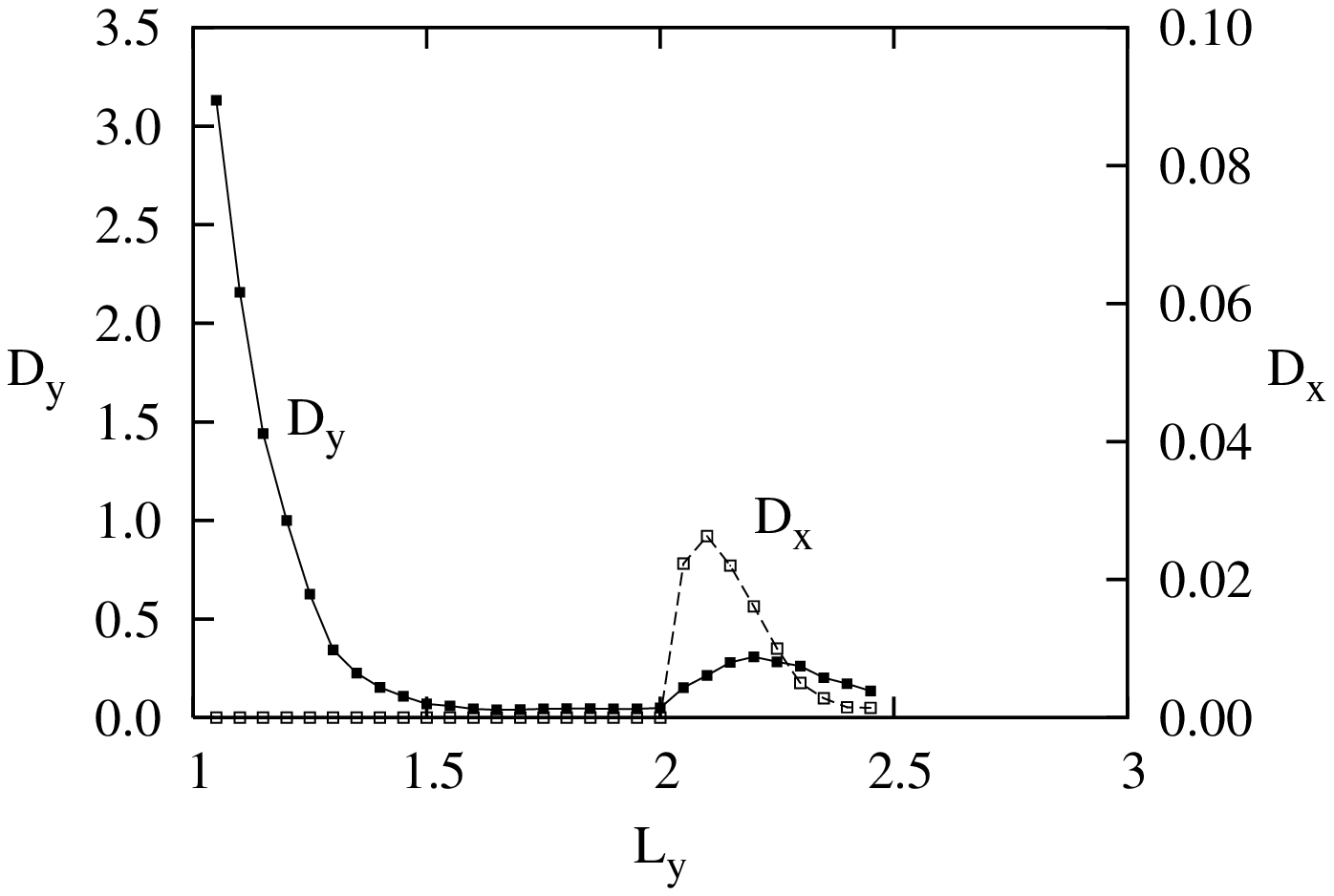}}
\caption{Effective diffusion constants (see the main text) in $x$ and
$y$ directions for disks in a narrow-channel with periodic boundaries
at a density $\rho = 0.8$.}
\label{diffusion}
\end{figure}
linear growth of $\langle \Delta x^2\rangle$ and $\langle \Delta y^2\rangle$
for times $t > 40$.  The results for a density $\rho = 0.8$ are shown in
Figure \ref{diffusion}. As expected $D_x$ vanishes for $L_y < 2$, while  $D_y$
is non-zero. In fact,  $D_y$ increases with $L_y$ approaching unity as a result
of the periodic-boundary restrictions disfavoring momentum exchange in the
$y$ direction. Another intersting feature of Fig. \ref{diffusion} is
the vanishing of $D_x$ and $D_y$ for $L_y > 2.7$. This is due
solid-like triangular configurations characteristic of the bulk at
this high density.

\section{Some remarks on Lyapunov spectra}
\label{spectra}
\begin{figure}
\centering{\includegraphics[width=8cm,angle=-90]{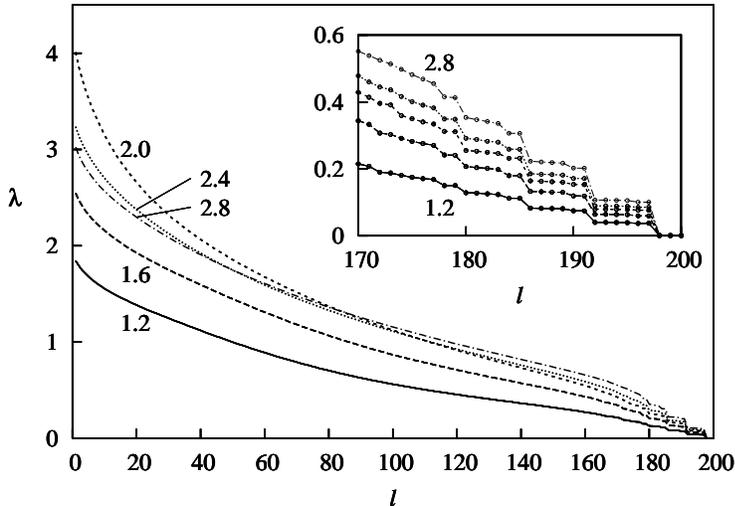}}
\caption{Positive branches of Lyapunov spectra for 100-disk
systems with {\em periodic} boundaries.  The box width, $L_y$, is
indicated by the labels. The density, $\rho = 0.4$, and the
Volume, $L_x L_y$, are fixed. The Lypunov exponents are defined
only for integer values of the index $l$. In the inset the regime
supporting Lyapunov modes is magnified. There, $L_y$ increases
from 1.2 (bottom) to 2.8 (top) in steps of 0.4. For details we
refer to the main text.} \label{spectrum}
\end{figure}

  In  Fig. \ref{spectrum} we plot the Lyapunov
spectra for a 100-disk system in a narrow periodic box.
All spectra are for a density $\rho = 0.4$, which
corresponds to a fluid. Since the volume, $L_x L_y$, is fixed,
an increase of the channel width $L_y$, varied in the figure
in steps of 0.4, decreases $L_x$ accordingly. $L_y$ is indicated by
the labels. Each spectrum consists of $4N$
Lyapunov exponents, of which only the positive branch,
$\{\lambda_l \ge 0; l = 1,2,\ldots,2N\}$, is shown. The spectra are
only defined for integer values of the index, $l$,
which labels the exponents according to size. The lines are
only drawn for clarity. According to the
equilibrium formulation of the conjugate-pairing rule, the negative
branch of the spectra, $\{\lambda_l \ge 0; l = 2N+1,\ldots,4N\}$, is
the mirror image of the positive branch \cite{pairing}.

   The inset of Fig. \ref{spectrum} provides a magnified view of the
small exponents. They turn out to be degenerate, which gives rise
to a step-like appearance of the spectra \cite{DPH96}. The
tangent-space perturbations associated with these exponents are
collective, wave-like fields defined over the simulation box, and
are referred to as Lyapunov modes \cite{PH00,FHPH04,FP03}. The
multiplicity  is determined by the intrinsic symmetries of the
Hamiltonian and the boundary conditions, which give rise to the
conserved quantities, energy and momentum. A complete
classification in terms of transverse (T), longitudinal (L), and
momentum (P) modes is given in Ref. \cite{ZEFP04}. Theoretical
attempts have been made to interpret the modes in terms of
fluctuating hydrodynamics \cite{EG00,MN01,MN03,TDP02,WB03}.

     We observe that a particular exponent, say the smallest
positive, increases with $L_y$ and, hence, with the wave
number, $k = 2 \pi/L_x = 2 \pi L_y/V$, belonging to this mode.
This is referred to as a ``dispersion relation''
\cite{PH00,FHPH04}. Thus, $\lambda_{2N-3}$ increases monotonously
with $L_y$. However, as Fig. \ref{spectrum} shows, this
proportionality does not hold for the large exponents. There is a
crossover of the spectra reversing the sequence of $\lambda$ near
$L_y = 2$. This is mainly a consequence of the collision frequency.
The perturbations for the large exponents are found to be localized
in space with only a small fraction of the particles contributing
to the large exponents at any instant of time. The active zone
moves around in space, restoring homogeneity on average.

\begin{figure}
\centering{\includegraphics[width=10cm]{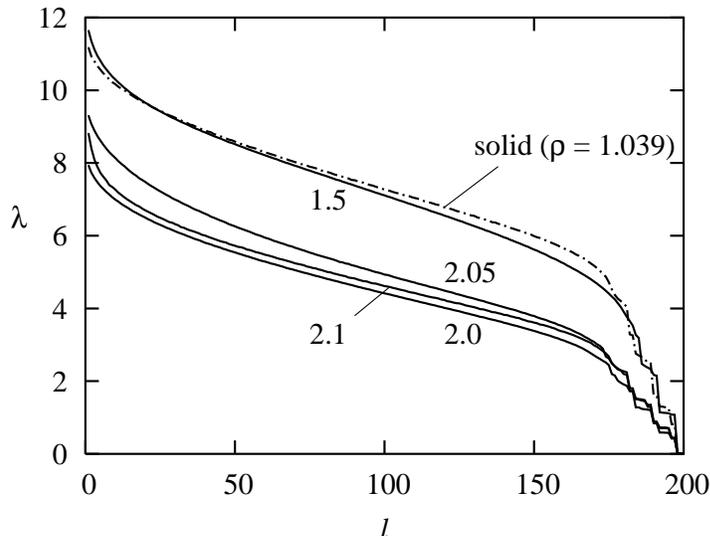}}
\caption{The Lyapunov spectra plotted by the smooth lines are for
100-disk systems with periodic boundaries, where the.
density, $\rho = 0.9$, and the volume, $L_x L_y$, are fixed.
The resulting box widths, $L_y$, are indicated by the labels.
The dash-dotted curve labeled ``solid'' is for 100 disks in a
periodic channel with an aspect ratio $L_y/L_x = \sqrt{3}/(N/2)$
required for a triangular lattice. The density $\rho$ is also given.
For details we refer to the main text.
Only the positive branches of the spectra are shown.}
\label{spectrum_0.9}
\end{figure}
     The smooth lines in Fig. \ref{spectrum_0.9} show
Lyapunov spectra for analogous 100-disk systems for a density
$\rho = 0.9$, where the
channel width is indicated by the labels. The case
$L_y = 1.5$ corresponds to the arrested structure of
Fig. \ref{structure_0.9} and is of particular interest. In
Fig. \ref{spectrum_0.9} we compare
it with the dash-dotted spectrum  of  $100$ disks in a
periodic channel with an aspect ratio $L_y/L_x = \sqrt{3}/(N/2)$.
For $L_y = 1.8258$ and $N=100$, the density is  $\rho = 1.039$, and the
particles form a triangular lattice characteristic of a solid in the
bulk. Although the nearest neighbor separations and, hence, the collision
frequencies, are nearly the same for both systems, the Lyapunov spectra are
distinctly different. The maximum exponent, $\lambda_1$, for the
arrested structure is significantly larger than for the triangular system,
whereas the behavior of the smaller exponents is just the reverse.
This example
demonstrates that the Lyapunov spectra are sensitive to structural
details and are potentially a useful tool. But it is fair to say that
no theory exists at present to interpret such spectra in full detail.

\section{Discussion}
\label{discussion}

     In this work the thermodynamic and
dynamical behaviour of hard disks in a narrow channel is analysed
using analytical and numerical approaches. The main results are
for periodic boundary conditions in both $x$ and $y$ directions.
Related studies of reflecting boundaries parallel to the channel
axis reveal that similar features exist there as well. It is found
that the pressure, collision frequency, maximum Lyapunov exponent,
and Kolmogorov-Sinai entropy curves exhibit a singularity for a
channel width equal to twice the disk diameter. For low densities
this singularity for $P$ and $\nu_2$ is well understood  by a
virial expansion as it already shows up in the second virial
coefficient. The singularity is not the result of a collective
behavior and, in fact, is present for systems with a finite number
of particles. Thus, it is not a genuine phase transition. It is
rather a result of a rapid change in the available phase space
taking place close to $L_y = 2$. Similar but less pronounced
singularities are expected to show up at $L_y = 3$ and larger
integers at higher virial coefficients.

     The singularity for the maximum exponent and for the KS-entropy
is most pronounced for higher densities. This is a consequence of the
fact that these quantities are closely related to the collision frequency
\cite{ZvBD}. It has been established that in bulk systems these quantities
 exhibit a maximum at a phase transition \cite{PHH90,DP97}. The maximum Lyapunov
exponent is a measure of the fastest dynamical events taking place in a system
which are localized processes in space \cite{FHPH04}. Thus, its behavior near
the transition point is not related to a diverging length scale
at the transition. Whereas in an ordinary phase transition in the bulk the
enhanced collision frequency - and, thus,  of $\lambda_1$ - is due to
the emergence of a new structure, here it is a consequence of the
constraints imposed by the boundaries.

    We expect the singularity found for $L_y = 2$ to be present also
for soft disks. In this case, however, the singularity is likely
to be weaker than the square-root singularity found in the
pressure curve of hard disks.  Similar features found in the
present study should be present in three-dimensional channels as
well. It would be of interest to study these problems in more
detail.

\acknowledgments

   We thank the Austrian Science Foundation (FWF),
Grant P15348-PHY, and the Israel Science Foundation for support.
We also acknowledge the Einstein Center and the
\"Osterreichische Gesellschaft der Freunde des Weizmann Institute of Science
for support during mutual visits.


\begin{thebibliography}{99}
\bibitem{Stillinger} T. M. Truskett, S. Torquato, S. Sastry,
     P. G. Debenedetti, and F. H. Stillinger,
     Phys. Rev. E {\bf 58}, 3083 (1998).
\bibitem{PH00} H. A. Posch and R. Hirschl, p. 269,
      in {\em Hard Ball Systems and the Lorenz Gas},
      edited by D. Szasz, Encyclopedia of the mathematical sciences
      {\bf 101}, Springer Verlag, Berlin (2000).
\bibitem{FHPH04} Ch. Forster, R. Hirschl, H. A. Posch, and Wm. G. Hoover,
      Physica D, {\bf 187}, 281 (2004).
\bibitem{DPH96} Ch. Dellago, H.A. Posch, and W.G.Hoover,
      Phys. Rev. E {\bf 53}, 1485 - 1501 (1996).
\bibitem{Woj82} K. W. Wojciechowski, P.Pieranski, and J. Malecki,
       J. Chem. Phys. {\bf 76}, 6170 (1982).
\bibitem{Woj83} K. W. Wojciechowski, P.Pieranski, and J. Malecki,
      J. Phys. A: Math. Gen.{\bf 16}, 2197 (1983).
\bibitem{VG03} S. Viscardy and P. Gaspard,
      Phys. Rev. E {\bf 68}, 041204 (2003).
\bibitem{Awazu} A. Awazu, Phys. Rev. E {\bf 63}, 032102 (2001).
\bibitem{Munakata} T. Munakata and G. Hu, Phys. Rev. E {\bf 65}, 066104 (2002).
\bibitem{HA67} W. G. Hoover and B. J. Alder, J. Chem. Phys.
        {\bf 46}, 686 (1967).
\bibitem{ZvBD} R. van Zon, H. van Beijeren, and J. R. Dorfman, p. 321,
      in {\em Hard Ball Systems and the Lorenz Gas},
      edited by D. Szasz, Encyclopedia of the mathematical sciences
      {\bf 101}, Springer Verlag, Berlin (2000).
\bibitem{Rapaport} D. C. Rapaport, {\em The Art of Molecular
      Dynamics Simulation}, Cambridge University Press, 2001, page 296.
\bibitem{AT} M. P. Allen and D. J. Tildesley, {\em Computer Simulation of
      Liquids}, Oxford University Press, Oxford (1991).
\bibitem{Oseledec} V. I. Oseledec, Trans. Moskow Math. Soc. {\bf 19},
      197 (1968).
\bibitem{AW70} B. J. Alder and T. E. Wainwright,
       Phys. Rev. A {\bf 1}, 18 (1970).
\bibitem{pairing} C. P. Dettmann and G. P. Morriss,
      Phys. Rev. E {\bf 53}, R5541 (1996).
\bibitem{FP03} Ch. Forster and H. A. Posch, Heareus Summer School
       Proceedings (2003).
\bibitem{PHH90} H.A. Posch, W.G. Hoover and B.L. Holian,
      Ber. d. Bunsenges. f. Phys. Chem.
     {\bf 94}, 250 - 256 (1990).
\bibitem{DP97} Ch. Dellago and H. A. Posch,
      Physica A, {\bf 237}, 95 - 112 (1997).
\bibitem{ZEFP04} E. Zabey, J.-P. Eckmann, Ch. Forster, and H. A. Posch,
       in preparation (2004).
\bibitem{EG00} J.-P. Eckmann and O. Gat, J. Stat. Phys. {\bf 98}, 775 (2000).
\bibitem{MN01} S. McNamara and M. Mareschal, Phys. Rev. E {\bf 64}, 051103
      (2001)
\bibitem{MN03} M. Mareschal and S. McNamara, Physica D {\bf 187}, 311 (2004).
\bibitem{TDP02} T. Taniguchi, C. P. Dettmann, and G. C. Morriss,
        J. Stat. Phys. {\bf 109}, 747 (2002).
\bibitem{WB03} A. de Wijn and H. van Beijeren, submitted to
        Phys. Rev. E  (2004).

\end{thebibliography}
\end{document}